\documentclass[reprint, amsmath, aps, pra, groupedaddress, superscriptaddress, longbibliography, floatfix, 
]{revtex4-2}
\usepackage[T1]{fontenc}
\usepackage{graphicx}
\usepackage{dcolumn}
\usepackage{bm}
\usepackage{amsmath}
\usepackage{amsfonts}
\usepackage{mathtools}
\usepackage[colorlinks,citecolor=black,linkcolor=black,urlcolor=black]{hyperref}
\usepackage{siunitx}
\usepackage[normalem]{ulem}
\usepackage{upgreek}
\usepackage{braket}
\usepackage{ragged2e}
\usepackage{subcaption}
\usepackage{mathtools}
\usepackage{xcolor}
\usepackage[format=plain,labelfont=bf,labelsep=space]{caption}
\DeclareCaptionJustification{mym}{\justifying}
\begin{document}

\title{Engineering Tunable Synthetic Su-Schrieffer-Heeger Chains in Liquid Crystal Microcavities}

\author{Joanna Mędrzycka}
\affiliation{Institute of Experimental Physics, Faculty of Physics, University of Warsaw, Poland}
\author{Luciano S. Ricco}
\affiliation{Institute of Experimental Physics, Faculty of Physics, University of Warsaw, Poland}
\author{Piotr Kapuściński}
\affiliation{Institute of Experimental Physics, Faculty of Physics, University of Warsaw, Poland}
\author{Marcin Muszyński}
\affiliation{Institute of Experimental Physics, Faculty of Physics, University of Warsaw, Poland}
\author{Przemysław Morawiak}
\affiliation{Institute of Applied Physics, Military University of Technology, Warsaw, Poland}
\author{Rafał Mazur}
\affiliation{Institute of Applied Physics, Military University of Technology, Warsaw, Poland}
\author{Rafał Węgłowski}
\affiliation{Institute of Applied Physics, Military University of Technology, Warsaw, Poland}
\author{Eva Oton}
\affiliation{Institute of Applied Physics, Military University of Technology, Warsaw, Poland}
\author{Przemysław Kula}
\affiliation{Institute of Chemistry, Military University of Technology, Warsaw, Poland}
\author{Wiktor Piecek}
\affiliation{Institute of Applied Physics, Military University of Technology, Warsaw, Poland}
\author{Jacek Szczytko}
\email{Jacek.Szczytko@fuw.edu.pl}
\affiliation{Institute of Experimental Physics, Faculty of Physics, University of Warsaw, Poland}
\date{May 2026}

\maketitle

\onecolumngrid
Optical microcavities have emerged as a powerful platform for emulating topological phases challenging to realize in conventional materials, offering precise control over dispersion, light confinement, and interactions. Among them, liquid crystal microcavities (LCMCs) offer exceptional tunability at room temperature, enabling voltage-controlled polarisation splitting, photonic spin-orbit coupling, and photonic potentials generated by self-assembled textures, such as cholesteric torons and uniform lying helix (ULH). Here, we design a LCMC hosting a dimerized ULH texture and show that the corresponding photonic potential describes two coupled Su-Schrieffer-Heeger chains with orthogonal linear polarisations, acting as an effective pseudospin degree of freedom. The applied voltage tunes the interchain coupling, enabling polarisation-dependent interactions. These results establish LCMCs as a versatile platform for tunable synthetic topological Hamiltonians.
\vspace{2em}

\twocolumngrid
\section{Introduction}

Topological phases of matter have become a central topic in solid-state physics across a wide range of platforms~\cite{Qi_RevModPhys.83.1057(2011),Lu_NatNanophot_2014}, from electronic systems to photonics, driven by both fundamental interest and application potential of robust, symmetry-protected phenomena~\cite{Bansil_RevModPhys.88.021004(2016),Chiu_RevModPhys.88.035005(2016)}. Representative examples include fault-tolerant quantum computing schemes based on topological superconductors~\cite{Nayak_RevModPhys.80.1083(2008),Book_topological_matter_quantum_computation}, as well as topological lasers realized in one-dimensional (1D) arrays of pillar microcavities~\cite{St_Jean_NatNanophot_2017}, exciton-polariton topological insulators~\cite{Klembt_Nature_2018}, and in time-reversal-symmetry-breaking two-dimensional (2D) photonic crystals~\cite{Ota_Nanophotonics_2020,Lan_RevPhys_2022}.

From a theoretical perspective, several minimal models have been introduced to describe and classify topological phase transitions. These include the Kane-Mele~\cite{Book_topological_insulators,Kane1_PhysRevLett.95.146802_2005,Kane2_PhysRevLett.95.226801_2005} and Bernevig-Hughes-Zhang (BHZ)~\cite{BHZ_Science.1133734_2006} models for 2D topological insulators as well as 1D models such as the Kitaev~\cite{A_Yu_Kitaev_2001} and Su-Schrieffer-Heeger (SSH) chains~\cite{SSH_PhysRevLett.42.1698_1979}, which capture topological superconductivity and topological insulating phases, respectively. The SSH model, in particular, is widely regarded as one of the simplest lattice systems behaving as a topological insulator, making it a common textbook example~\cite{Book_topological_insulators}. It simply consists of a dimerized tight-binding chain with two sublattices per unit cell, characterized by alternating intracell and intercell nearest-neighbour hopping amplitudes. The ratio between these couplings determines the topological phase of the system.

Conceptually, the SSH model can be viewed as the simplest limit of a broader class of 1D topological Hamiltonians. Extensions have been proposed to include coupling between stacked chains~\cite{Agrawal_PhysRevB.108.104101(2023)}, next-nearest-neighbour (NNN) hoppings~\cite{Linhu_PhysRevB.89.085111(2014),Jiao_PhysRevLett.127.147401(2021),DiSalvo_PhysRevB.110.165145(2024)}, spin-orbit coupling (SOC)~\cite{Kokhanchik_PhysRevLett.129.246801(2022),Whittaker_PhysRevB.99.081402_2019,Yan_EPL_10747007(2014),Bahari_PhysRevB.94.125119(2016),LiuPhysRevB.104.085302_2021}, superconducting pairing~\cite{rufo2025domainwallcontroltopological,Zhao_PhysRevB.110.235106_2024,Tamura_PhysRevB.101.214507_2020}, and Hubbard-type interactions~\cite{Zhao_PhysRevB.110.235106_2024,Chang_SciRep_2025,wang2026defectengineeringspincenters}. These additional terms give rise to a richer topological phase diagram, in which the transition between trivial and nontrivial phases is no longer governed solely by the ratio of intercell and intracell hopping amplitudes. In such modified SSH-like models, even when chiral symmetry is broken, inversion symmetry can protect nontrivial crystalline topological regimes characterized by a quantized bulk Zak phase~\cite{Longhi_OptLett_43_2018,Jiao_PhysRevLett.127.147401(2021),TaoDu_PhysicaE_2025,Zak_PhysRevLett.62.2747_1989,Delplace_PhysRevB.84.195452_2011}. As a result, the standard Altland-Zirnbauer (AZ) tenfold classification and its bulk-boundary correspondence are incomplete~\cite{Chiu_RevModPhys.88.035005(2016),Ryu_NewJourPhys_2010}, since topological phases may exist without symmetry-pinned boundary zero modes~\cite{Delplace_PhysRevB.84.195452_2011}.

Beyond its seminal formulation to describe solitons in polyacetylene~\cite{SSH_PhysRevLett.42.1698_1979}, SSH chains have been realized across a variety of experimental platforms, including self-assembly atomic chains~\cite{Jalochowski_ACSNano_2024}, trapped atoms~\cite{Meier_NatComm_2016}, quantum-dot chains~\cite{Pham_PhysRevB.105.125418_2022}, exciton-polariton lattices of micropillars~\cite{St_Jean_NatNanophot_2017,Whittaker_PhysRevB.99.081402_2019}, microcavity traps~\cite{Tristan_ACS_Photonics_2021}, and arrays of coupled waveguides~\cite{Cheng_LaserPhotRev_2015}. In most of these implementations, the lattice geometry and coupling strengths are defined during fabrication or setup and remain largely static, allowing only limited in-situ tunability. While these platforms have been instrumental in establishing and probing topological phases, they offer limited flexibility for exploring on-demand additional internal degrees of freedom, inter-chain couplings, or selective symmetry breaking.

Among experimental systems capable of realizing topological Hamiltonians, liquid crystal microcavities (LCMCs) offer a highly versatile platform for implementing synthetic photonic Hamiltonians with high tunability via, e.g., applied electric field~\cite{synthetic_Hamiltonians,spin_hall_in_lcmc}. In these systems, the intrinsic birefringence of the liquid crystal material gives rise to polarisation-dependent cavity modes, which in turn lead to an effective photonic SOC~\cite{synthetic_Hamiltonians,Kokhanchik_PhysRevLett.129.246801(2022)}, as well as polarisation textures ~\cite{merons, psh}, singularities~\cite{pol_singularities_lcmc} and ground-state lasing with induced optical angular momentum \cite{torony}. Furthermore, when the liquid crystal forms a self-organised uniform lying helix (ULH) structure~\cite{ulh_morphology}, it generates a spatially modulated photonic potential across the cavity plane~\cite{Muszyński_LasPhotRev2025}. This potential can be interpreted as a synthetic 1D lattice for photons, in which the modulation depth, polarisation-dependent hopping, and intersite coupling are all controllable via an applied voltage, providing an alternative to a deterministic substrate patterning~\cite{gaussian_traps}.

In this work, we demonstrate that LCMCs hosting a self-assembled ULH potential can realise a coupled double SSH chain, with the two polarisation-dependent SSH lattices interacting with each other. The applied external voltage controls the coupling between the chains. Further theoretical analysis shows that both chiral and inversion symmetries are broken for the entire voltage range explored here, preventing the emergence of a topological phase protected by crystalline inversion symmetry. Nevertheless, our analysis demonstrates that inversion symmetry protection can, in principle, be restored by tuning the system's parameters, allowing the existence of a topological region in parameter space. These findings establish LCMCs as a versatile and electrically reconfigurable platform for engineering beyond-minimal SSH Hamiltonians and exploring extended 1D topological regimes in optics, opening new avenues for on-demand simulation of synthetic lattices and more exotic topological phenomena.

\section{Results and Discussion}
\begin{figure*}[t]
    \centering
    \includegraphics[width=17cm]{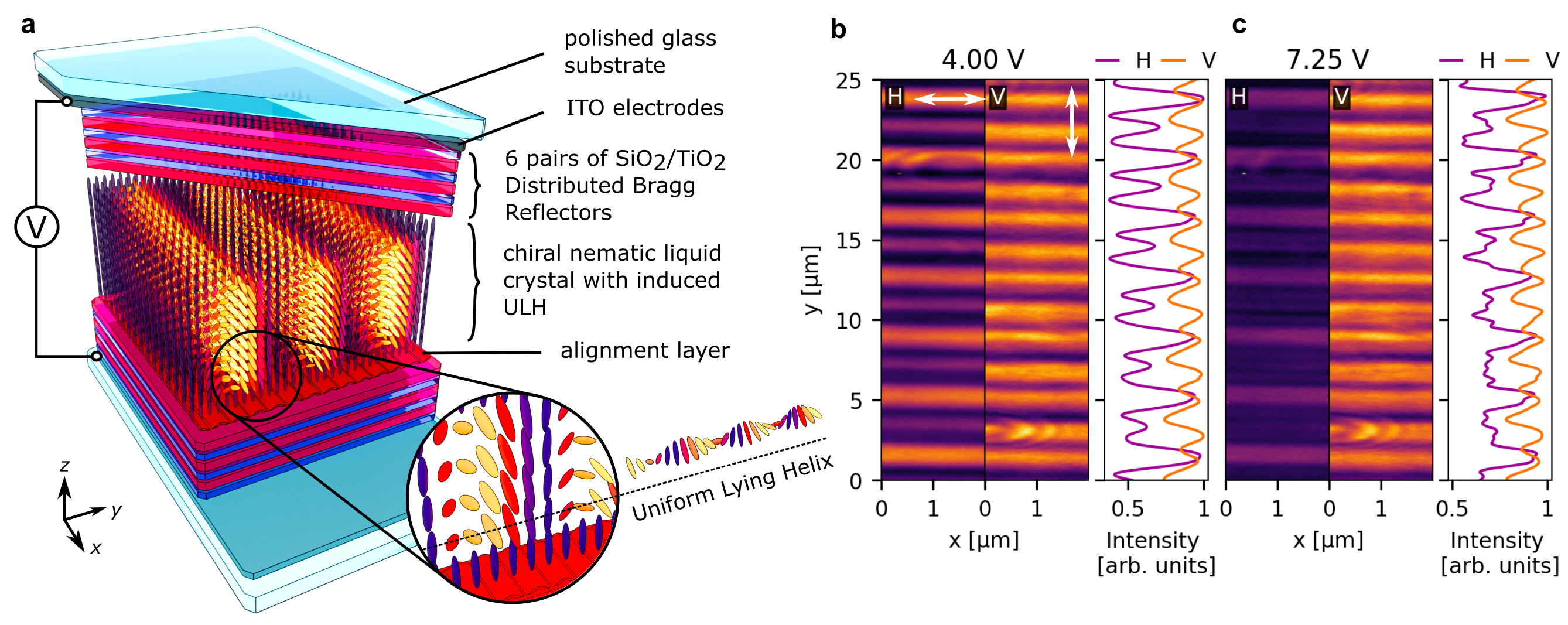}
    \caption{The ULH configuration inside an optical microcavity. A scheme of the measured sample (a), with $n_{x/y}$ and $n_z$ indicating the orientation of the liquid crystal director in the $xy$ and $zx/zy$ planes. White light transmission captures of the ULH texture in the H $\parallel x$ polarisation and V $\parallel y$  polarisation for the lowest (b) and highest (c) applied voltages, with corresponding side panels showing the integrated transmission signal along the transverse direction, exhibiting an alternating intensity pattern. Arrows in (b) indicate the H/V polarisation orientations corresponding to the $x/y$ axes. The microscope images were coloured for clarity.}
    \label{fig1}
\end{figure*}

Our platform is an LCMC sample formed by assembling two Distributed Bragg Reflectors (DBRs), coated with a homeotropically orienting polyimide layer to ensure proper liquid crystal alignment, and defining an initially air-filled microcavity. The cavity is subsequently filled with a liquid crystal mixture with a chiral dopant via capillary forces (see sample fabrication details in the supplementary material). In our case, the system operates in the uniform lying helix (ULH) configuration, which is formed and electrically controlled by an external voltage between 4.00~V and 7.25~V, applied using transparent Indium Tin Oxide (ITO) electrodes deposited on the rear sides of the DBR mirrors, as shown in Fig.~\ref{fig1}a. Outside of this voltage range, the liquid crystal unravels into a simple nematic ordering, achieved initially via a thermally driven phase transition. Due to the birefringence of the liquid crystal material, the ULH texture creates a spatially varying photonic potential which resembles a 1D dimerized lattice for both the horizontal (H) and vertical (V) polarisations, which lay in the transverse and longitudinal lattice directions, respectively. Importantly, this photonic potential dimerisation manifests directly in the spatial dependence of the cavity transmission intensity. Fig.~\ref{fig1}b presents a camera capture of the white light transmission measurement in both the H and V polarisation directions for the lowest applied voltage of 4.00~V. The transmission signal was subsequently integrated along the transverse direction, revealing an alternating intensity pattern characteristic of lattice dimerisation, with a more pronounced contrast for the H polarisation. This controlled dimerisation distinguishes the present work from our previous study, where a basic 1D ULH lattice without alternating site modulation was also realised in an LCMC~\cite{Muszyński_LasPhotRev2025}. Details about transmission spectra measurements are described in the supplementary material. 

These dimerised, polarisation-dependent potentials are dynamically tunable via the electro-optical response of the liquid crystal. The external field reorients the liquid crystal molecules, with the reorientation initially occurring for molecules oriented closest to the field direction, thereby modifying the effective refractive index anisotropy and consequently reshaping the photonic potential landscape. Fig.~\ref{fig1}c depicts the camera image of the white light transmission measurement in the H and V polarisations for the maximum applied voltage of 7.25~V. This liquid crystal reorientation is observed as a pronounced spatial shift in the transmitted intensity for H polarisation, whereas the V polarisation remains largely unchanged. This behaviour is consistent with a field-induced twist of the liquid crystal director predominantly within the horizontal plane. At a high electric field, the H-polarised potential preserves its dimerised structure, although it exhibits reduced spatial contrast.

\begin{figure*}[t]
    \centering
    \includegraphics[width=17cm]{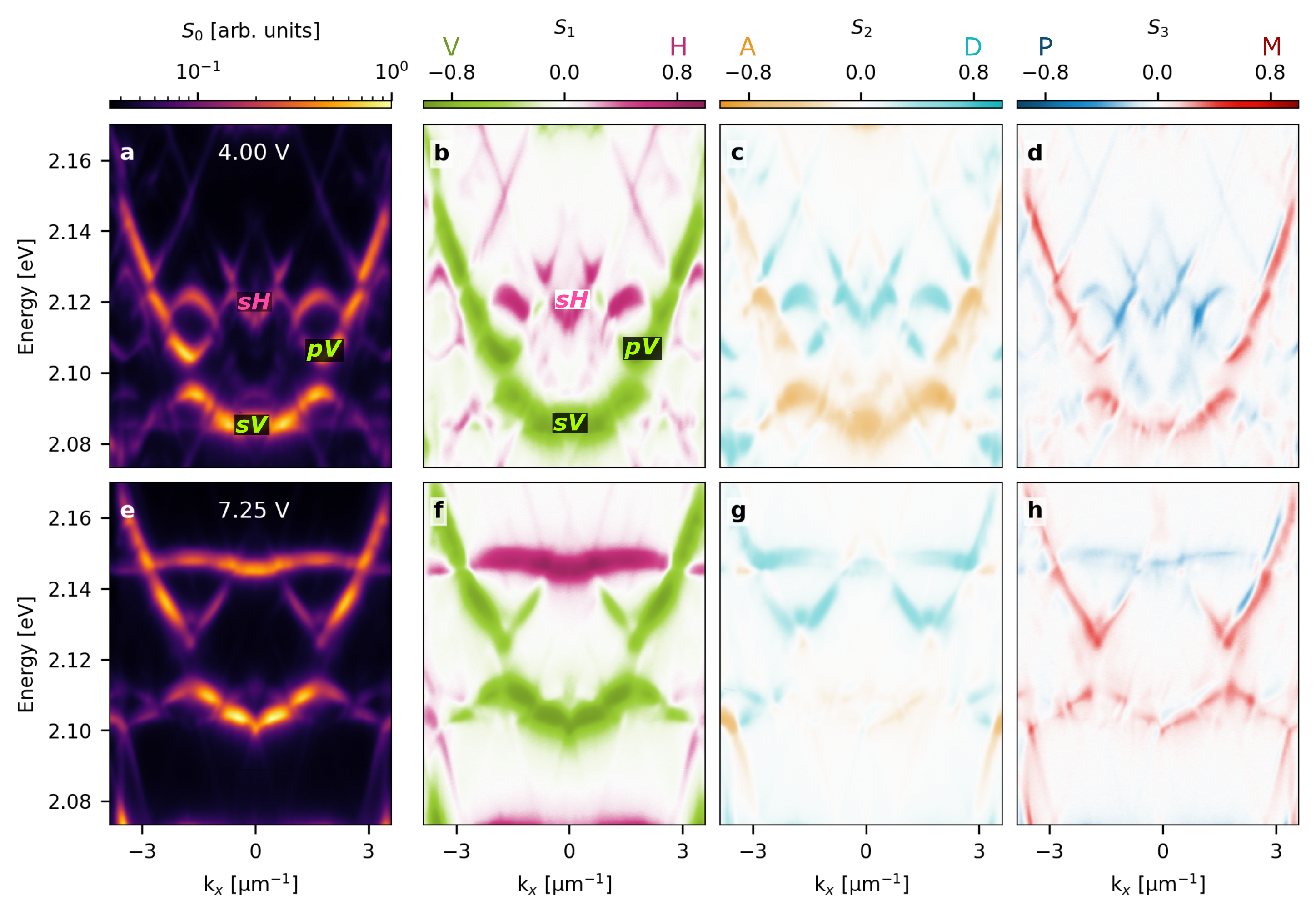}
    \caption{The reciprocal-space transmission spectra for the lowest (top row) and highest (bottom row) voltages applied to the sample. The $S_0$ Stokes parameters (a, e) shown with a logarithmic colour scale. The $S_1$ (b, f), $S_2$ (c, g) and $S_3$ (d, h) Stokes parameters, respectively, exhibit the polarisation degree of the energy bands. Band labels introduced in the main text were added to (a) and (b) for convenience.}
    \label{fig2}
\end{figure*}

The ULH-induced 1D photonic potential opens bandgaps in the cavity modes~\cite{Muszyński_LasPhotRev2025}, as depicted in Fig.~\ref{fig2}. Fig.~\ref{fig2}a and \ref{fig2}e present the normalised intensity of the transmitted light for the lowest and highest applied voltages, revealing three distinct photonic bands within the otherwise parabolic dispersion of microcavity photons not bound by any additional photonic potential. Furthermore, analysis of the Stokes parameters shows distinct polarisation characteristics for each photonic band, as depicted in Fig.~\ref{fig2}b-d and \ref{fig2}f-h.

The Fabry–Pérot optical modes in the cavity exhibit a parabolic dispersion relation $E(\vec{k})$ \cite{kavokin2007microcavities}. The presence of a periodic potential leads, to first approximation, to a replication of each mode in k-space, shifted by reciprocal lattice vectors (in our case it is $\mathrm{k}_x\approx\pm1.7$~$\upmu\mathrm{m}^{-1}$). This is reflected in the appearance of successive intersecting branches in the energy-momentum dispersion $E(\vec{k})$.
In the absence of interactions between the modes, the dispersion resembles that of the well-known \textit{nearly free electron model} in solid-state physics: identical parabolic bands shifted in momentum space, intersecting without hybridization, corresponding to free and non-interacting quasiparticles in a periodic potential \cite{IbachLuch}. However, when coupling between the states is taken into account, energy gaps open at the intersection points, and the spectrum reorganizes into energy bands. The strength of the interaction determines the size of these gaps.

The lowest band, centred around 2.09~eV (Fig.~\ref{fig2}), exhibits an approximately sinusoidal dispersion. Additional anti-crossings are observed, likely resulting from interaction with dispersions of lower cavity modes. The band shows a sinusoidal bright state and a corresponding dark state, consistent with the characteristic bonding-antibonding structure of an SSH lattice and indicating minimal interaction with other bands. Analysis of the Stokes parameters (Fig.~\ref{fig2}b-d) shows that the band is predominantly vertically polarised, hence denoted as the \textit{sV} band for simplicity. The $S_2$ Stokes parameter (Fig.~\ref{fig2}c) indicates a primarily anti-diagonal (AD) polarisation component of the \textit{sV} band. This anti-diagonal component disappears at the higher applied voltages (Fig.~\ref{fig2}g), reflecting a voltage-dependent modification of the inter-band interactions.
The $S_3$ parameter shows no evidence of the chequerboard polarisation pattern characteristic of the inter-subband spin-orbit coupling (ISOC). This contrasts with the 1D ULH lattice studied previously, which exhibited clear ISOC-mixed states and corresponding optical activity~\cite{Muszyński_LasPhotRev2025}.

The behaviour of the higher bands centred around 2.11~eV proves more complex (see Fig.~\ref{fig2}a), as they depict a non-trivial dispersion far from a sinusoidal shape. Following a standard ordering of the energy states, above the so-called \textit{sV} state, we should expect to see a predominantly V polarised \textit{p}-like state or a predominantly H polarised \textit{s}-like state. The H and V polarisation components, however, are strongly mixed, with visible discontinuities, particularly for the V polarisation around~$\mathrm{k}_x=\pm1$~$\upmu\mathrm{m}^{-1}$ in Fig.~\ref{fig2}b. Such shape indicates a significant coupling between the \textit{s}-like and \textit{p}-like bands of opposite polarisations (pseudospins). Additionally, the mixing of bonding and anti-bonding states of both modes further contributes to this behaviour. The resulting band in the next V polarised state with a \textit{p}-like shape is labelled as the \textit{pV} band, while the upper, predominantly H polarised component is called the \textit{sH} band to simplify further discussion. Additional bandgaps are also visible in these bands, particularly at 2.11~eV for the \textit{pV} band or at 2.12~eV for the \textit{sH} band. Above 2.13~eV, the photon modes become unbound, returning to a parabolic energy dispersion. 

The energy splitting between the ground states of microcavity modes in the H and V polarisations stems from the birefringence of the liquid crystal medium \cite{synthetic_Hamiltonians, spin_hall_in_lcmc}. Importantly, this birefringence leads to the degeneracy of the \textit{s}-like and \textit{p}-like bands, enabling an effective coupling between them and leading to the unique shapes of the \textit{pV} and \textit{sH} bands. In contrast, there is no indication of significant interaction between the \textit{sV} band and any of the higher bands for the voltage range considered, and hence, it shall be excluded from further theoretical analysis.

The liquid crystal’s response to the applied electric field strongly modifies the reciprocal-space transmission spectra. As shown by comparison between the top and bottom panels of Fig.~\ref{fig2}, all photonic bands shift to higher energies, with noticeable changes in their dispersion shapes. This effect is particularly pronounced in the case of the \textit{sH} band, which evolves from a non-trivial shape at 4.0~V (Fig.~\ref{fig2}e) towards a near-flat band at 7.25~V (Fig.~\ref{fig2}f). indicating an increase in the photonic effective mass. The \textit{pV} band also changes its shape, showing a closure of the bandgaps previously seen for the lowest applied voltage around 2.11~eV or 2.12~eV. Additionally, the $S_2$ Stokes parameter reveals voltage-dependent polarisation changes in the \textit{pV} and \textit{sH} bands. For the lowest applied voltage (Fig.~\ref{fig2}c), the degree of diagonal polarisation is mixed, and shifts from an anti-diagonal (AD) to a diagonal (D) around the first Brillouin Zone (BZ) edges ($\mathrm{k}_x\approx\pm1.7$~$\upmu\mathrm{m}^{-1}$). At higher voltage, the polarisation of both bands changes from a mixed D-AD to a primarily D component, indicating that the $sH$-$pV$ interactions are strongly voltage-dependent. The $S_3$ Stokes parameter remains negligible, confirming the absence of a significant degree of circular polarisation.

\begin{figure*}[t]
    \centering
    \includegraphics[width=17cm]{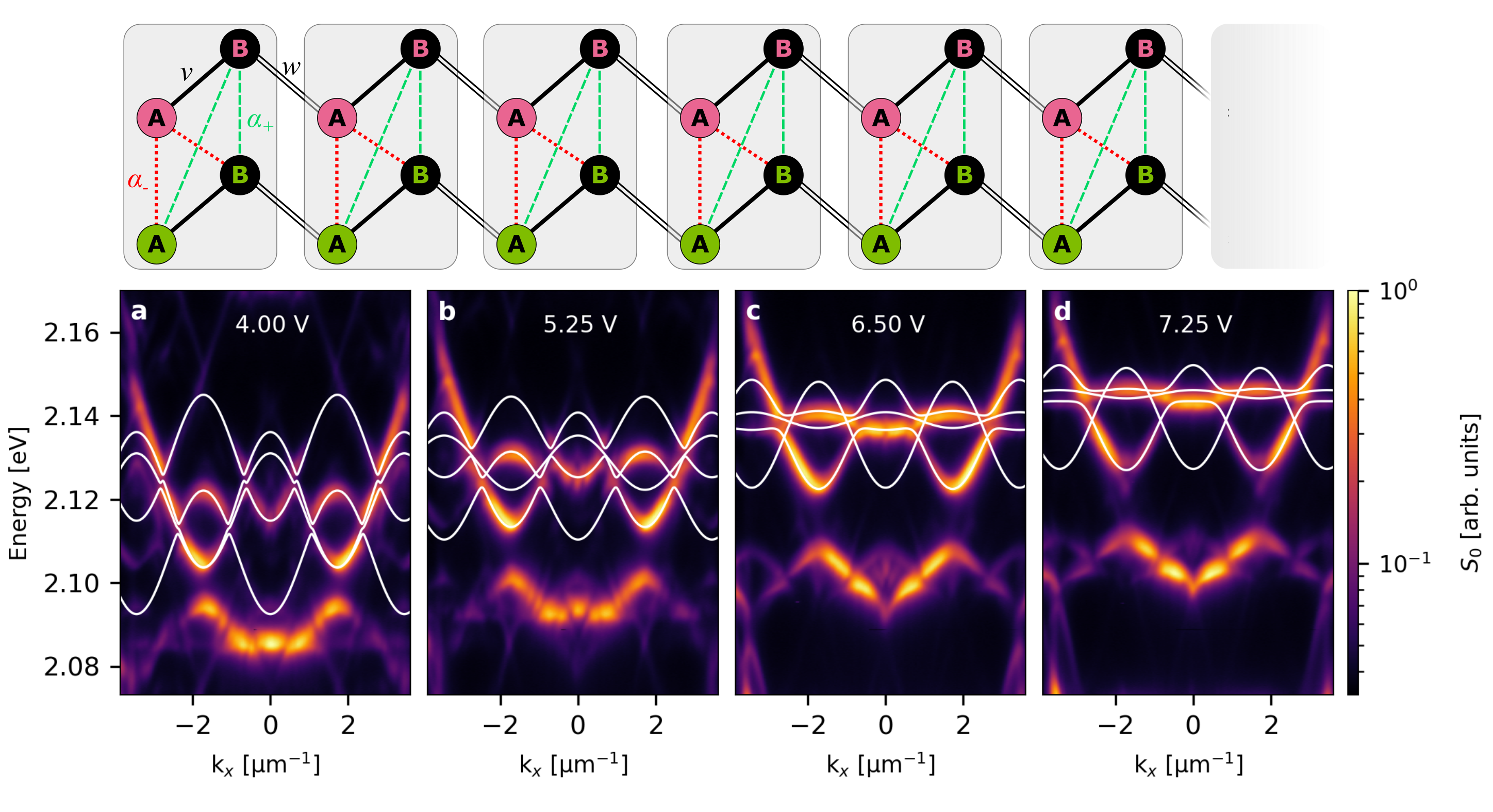}
    \caption{Top panel exhibits the sketch of the two coupled SSH chains with staggered onsite potentials, and intracell and intercell hoppings $v$ (single black line) and $w$ (double black line), respectively, with interchain hopping amplitudes $\alpha_-$ (dotted red line) and $\alpha_+$ (dashed teal line). Each chain has a distinct orbital-polarisation, denoted as $sH$ (upper chain) and $pV$ (lower chain). (a-d) Reciprocal-space transmission spectra $S_0$ in logarithmic scale, considering four values of applied voltage. The overlaid white lines represent the fitting using the effective reciprocal-space Hamiltonian given by Eq.~\eqref{H(k)_matrix}, with the parameters shown in Table S1 of the supplementary material. 
    }
    \label{fig3}
\end{figure*}

The behaviour of the relevant \textit{pV} and \textit{sH} bands for distinct applied voltages shown in Fig.~\ref{fig2}, stemming from the tunable effective 1D dimerised photonic potential (Fig.~\ref{fig1}b,c), suggests that our system can be described phenomenologically within the SSH model framework. Particularly, we consider two coupled SSH chains~\cite{Book_topological_insulators,SSH_PhysRevLett.42.1698_1979} with staggered onsite potentials and opposite \textit{pV-sH} orbital-polarisations. The model parameters, including the coupling between the SSH chains, are modulated via the applied electric field through voltage change. Namely, such a field modifies the mean refractive index anisotropy of the ULH texture and hence, the effective photonic potential. Fig.~\ref{fig3}, top panel, depicts the scheme of the two coupled SSH chains mimicking the corresponding photonic potential landscape, which is captured by the following effective Hamiltonian:
\begin{equation}
    \hat{H} = \hat{H}_{sH} + \hat{H}_{pV} + \hat{H}_{\text{mix}},\label{eq:H_real_space}
\end{equation}
where 
\begin{eqnarray}
    \hat{H}_{\sigma} & = &\sum_n \left[(E_0^{\sigma}-\epsilon_\sigma) \hat{a}_{n,\sigma}^\dagger \hat{a}_{n,\sigma} + (E_0^{\sigma}+\epsilon_\sigma)\hat{b}_{n,\sigma}^\dagger \hat{b}_{n,\sigma}  \right] \nonumber \\
    &+& \left(v_{\sigma}\sum_{n}^{N}\hat{a}_{n,\sigma}^\dagger \hat{b}_{n,\sigma} + w_{\sigma}\sum_{n}^{N-1}\hat{a}_{n,\sigma}^\dagger \hat{b}_{n+1,\sigma} + \text{h.c.} \right),\label{eq:H_sigma_Rspace}
\end{eqnarray}
describes a single SSH chain of $N$ unit cells, each of them having two sites represented by sublattices A and B, with orbital-polarisation $\sigma=sH,pV$. The operators $\hat{a}_{n,\sigma}^\dagger$ and $\hat{b}_{n,\sigma}^\dagger$ create a particle with (pseudo) spin $1/2$ in the sublattices A and B, respectively, at unit cell $n$ with orbital-polarisation (pseudospin) $\sigma$, with $E_{0}^\sigma$ being the onsite energy and $\pm\epsilon_\sigma$ representing the energy difference between the sublattice sites of each unit cell, giving rise to the staggered potential pattern. The parameters $v_\sigma$ and $w_\sigma$ are the intercell and intracell hopping amplitudes, respectively, and h.c. stands for the hermitian conjugate. Notice that for $E_0^\sigma = 0$, Eq.~\eqref{eq:H_sigma_Rspace} corresponds to the Hamiltonian for the Rice-Mele model~\cite{Book_topological_insulators}.

The chains of opposite orbital-polarisation are coupled to each other according to
\begin{eqnarray}
 \hat{H}_{\text{mix}} & = & \sum_n \left(\alpha_- \hat{a}_{n,sH}^\dagger\hat{a}_{n,pV} + \alpha_+ \hat{b}_{n,sH}^\dagger\hat{b}_{n,pV}  \right. \nonumber \\
  & - & \left. \alpha_+ \hat{a}_{n,sH}^\dagger\hat{b}_{n,pV} - \alpha_- \hat{b}_{n,sH}^\dagger\hat{a}_{n,pV} \right) + \text{h.c.},\label{H_mix}
\end{eqnarray}
where $\alpha_{\pm}=(\alpha_1 \pm \alpha_2)/2$ are the coupling amplitudes between the chains, with $\alpha_1$ and $\alpha_2$ being parameters to be fitted from experimental data (see the supplementary material for details).

The double SSH chain with staggered onsite potential described by Eq.~\eqref{eq:H_real_space} can be rewritten in reciprocal space by a standard discrete Fourier transform of sublattice operators $\hat{a}_{n,\sigma}$ and $\hat{b}_{n,\sigma}$ using plane waves. Following this procedure, the bulk momentum-space Hamiltonian reads:
\begin{equation}
    \hat{H}(k) = \boldsymbol{\Psi}_k^\dagger\boldsymbol{H}(k)\boldsymbol{\Psi}_k,\label{H(k)_full}
\end{equation}
with $\boldsymbol{\Psi}_k=(\hat{a}_{k,sH},\hat{b}_{k,sH},\hat{a}_{k,pV},\hat{b}_{n,pV})^T$, and the $4\times4$ Bloch Hamiltonian
\begin{equation}
 \boldsymbol{H}(k) = \begin{bmatrix}E_0^{sH}-\epsilon_{sH} & t_{sH}(k) & \alpha_{-}&-\alpha_+\\
 t_{sH}^*(k) & E_0^{sH} + \epsilon_{sH} & -\alpha_- &\alpha_+\\
  \alpha_{-} & -\alpha_- & E_0^{pV} - \epsilon_{pV} & t_{pV}(k)\\
   -\alpha_+ & \alpha_+ & t_{pV}^*(k)&E_0^{pV} + \epsilon_{pV}\\
\end{bmatrix}. \label{H(k)_matrix}  
\end{equation}
Here, $t_\sigma(k) = v_\sigma + w_\sigma e^{ikd}$, where $d$ is the lattice constant, i.e., the unit-cell spacing of each chain. By considering the approximation of a small topological gap, where $w_\sigma \rightarrow v_\sigma + \delta_\sigma$, with $\delta_\sigma \ll v_\sigma, w_\sigma$, the $k$-dependent hopping between A and B sublattices becomes $t_\sigma(k)\approx 2t_\sigma e^{\frac{ikd}{2}}\cos\left(\frac{kd}{2}\right)$. The overall phase prefactor $e^{\frac{ikd}{2}}$, corresponding to a unit-cell gauge choice, can be absorbed into the definition of the \textit{k}-space operators, allowing one to equivalently take $t_\sigma(k)\approx 2t_\sigma \cos\left(\frac{kd}{2}\right)$, with $t_\sigma = \sqrt{v_\sigma w_\sigma}$.

From Eqs.~\eqref{H_mix}–\eqref{H(k)_full}, the terms proportional to $\alpha_{\pm}$ couple opposite orbital-polarisation states (\textit{sH}-\textit{pV}) within the same unit cell (Fig.~\ref{fig3}, top), hybridising the otherwise independent SSH chains. Interpreting polarisation as a pseudospin, these terms represent a local inter-chain coupling between states of opposite spin and orbital parity. Although assumed momentum-independent within our effective model and thus not a genuine SOC, these couplings effectively describe sublattice-assisted spin-orbital mixing between opposite-parity bands. The voltage-dependent mixing observed in $S_2$ (Fig.~\ref{fig2}c,g) confirms such a controlled spin-orbital coupling, while the absence of optical activity in the $S_3$ (Fig.~\ref{fig2}d,h) rules out intrinsic SOC~\cite{Muszyński_LasPhotRev2025}.

By fitting $t_\sigma$, as well as the other parameters described by the Hamiltonian of Eq.~\eqref{H(k)_matrix}, we can reproduce the $k$-space transmission spectra given by $S_0$ for distinct values of applied voltages, as shown by the white curves overlaid on transmission profiles shown in the bottom panels of Fig.~\ref{fig3}. Data about fitting values are listed in Table.~S1 of the supplementary material. While other bands appear in the transmission spectra, our theoretical analysis is restricted to the bands that vary most significantly with increasing applied voltage.

Although the fitting does not provide independent values of the intercell ($w_\sigma$) and intracell ($v_\sigma$) hoppings for each decoupled chain and hence, does not fully determine their individual topology~\cite{Book_topological_insulators}, the gapped regions visible in Fig.~\ref{fig3}a,d support an analysis of the system’s symmetries and underlying topology in the presence of the interchain coupling $\alpha_\pm$~\cite{Chiu_RevModPhys.88.035005(2016)}. Previous studies have predicted distinct topological phases in SSH chains with two spin degrees of freedom under SOC-like modulation, even when the uncoupled chains are topologically trivial~\cite{Yan_EPL_10747007(2014),Bahari_PhysRevB.94.125119(2016)}. In LCMC systems, Rashba-Dresselhaus SOC has been shown to act as a synthetic gauge field, whose modulation can drive a spinless, topologically trivial SSH chain into a nontrivial phase~\cite{Kokhanchik_PhysRevLett.129.246801(2022)}.

To investigate the topological transitions, we analyse the chiral and inversion symmetries of the bulk Hamiltonian in Eq.~\eqref{H(k)_matrix}~\cite{Chiu_RevModPhys.88.035005(2016),Jiao_PhysRevLett.127.147401(2021),TaoDu_PhysicaE_2025} (see Supplemental Material for details). The staggered potential $\epsilon_\sigma$ breaks both symmetries even for a single SSH chain ($\alpha_\pm=0$), while interchain coupling breaks chiral symmetry even when $\epsilon_\sigma=0$. In contrast, inversion symmetry remains preserved when $\alpha_+=\alpha_-$ and $E_0^{sH}=E_0^{pV}$. Since chiral symmetry breaking prevents the definition of a winding number within the standard AZ classification~\cite{Chiu_RevModPhys.88.035005(2016),Jiao_PhysRevLett.127.147401(2021)}, topologically protected zero-energy edge modes are no longer guaranteed. Nevertheless, inversion symmetry alone can support crystalline topological phases, as in the representative case of a SSH chain with broken chiral symmetry~\cite{Longhi_OptLett_43_2018,Jiao_PhysRevLett.127.147401(2021),TaoDu_PhysicaE_2025}. In this situation, the Zak phase is the relevant topological invariant~\cite{Zak_PhysRevLett.62.2747_1989,Delplace_PhysRevB.84.195452_2011}. Its quantisation indicates a nontrivial bulk phase, although it does not determine the number of edge states, reflecting the limitations of bulk-boundary correspondence beyond the AZ framework.

In our experiment, for all fitting parameters corresponding to Fig.~\ref{fig3}, both chiral and inversion symmetries are broken, thereby placing the synthetic coupled SSH chains in a trivial phase. Inversion symmetry, however, can be restored when $\epsilon_\sigma=0$ and $\alpha_+=\alpha_-$, potentially yielding a quantized Zak phase and a topological regime driven by interchain coupling between opposite spin-parity states. An alternative route to break chiral symmetry while preserving inversion symmetry in SSH lattices is the inclusion of next-nearest-neighbour (NNN) couplings~\cite{Linhu_PhysRevB.89.085111(2014),Jiao_PhysRevLett.127.147401(2021),Agrawal_PhysRevB.108.104101(2023),DiSalvo_PhysRevB.110.165145(2024)}, corresponding to hybridization between identical sublattices of adjacent unit cells, $\hat{a}_n(\hat{b}_n)\leftrightarrow \hat{a}_{n+1}(\hat{b}_{n+1})$, either within the same or between distinct SSH chains.

\section{Conclusion}

We explored the high tunability of LCMCs to experimentally realize two coupled Su-Schrieffer-Heeger (SSH) chains with opposite polarisation and orbital parity. A self-assembled ULH texture creates an effective 1D dimerised, polarisation-dependent, lattice potential, controlled by an external voltage, and responsible for tuning the interchain coupling. The analysis of the photonic band structure shows that both chiral and inversion symmetries are broken across the explored voltage range, placing the system in a topologically trivial phase. Nevertheless, our theoretical investigation identifies a regime where inversion symmetry can be restored, enabling a topological region characterized by a quantized Zak phase, despite broken chiral symmetry. The intrinsic tunability of LCMCs further allows engineering next-nearest-neighbour intra and interchain hoppings, for instance, via nematic or cholesteric liquid crystal structures embedded in microcavities with pre-designed photonic potentials, patterned through a focused ion beam and combined with electric-field control. Overall, our results establish LCMCs as a versatile platform for synthetic topological Hamiltonians, including those supporting fractionally charged fermionic quasiparticles with non-Abelian braiding statistics\cite{Bahari_PhysRevB.94.125119(2016),Klinovaja_PhysRevLett.110.126402(2013),Nayak_RevModPhys.80.1083(2008)}, with on-demand tunability.

\section*{Funding}
Polish National Science Centre, Project No 2023/51/B/ST3/03025.
Polish National Science Centre, Project No 2021/43/I/ST5/02632 (HoloBlue). 

\section*{Disclosures}
The authors declare no conflicts of interest.

\section*{Data Availability Statement}
Data underlying the results presented in this paper may be obtained from the authors upon reasonable request.

\section*{Supplemental document}
See Supplement for supporting content.

\newpage
\label{sec:refs}
\bibliography{sample}

@article{spin_hall_in_lcmc,
author = {Rechcińska, Katarzyna and Król, Mateusz and Mirek, Rafał and Łempicka-Mirek, Karolina and Stephan, Daniel and Mazur, Rafal and Morawiak, Przemysław and Kula, Przemyslaw and Piecek, Wiktor and Lagoudakis, Pavlos and Pietka, Barbara and Szczytko, Jacek},
year = {2018},
month = {12},
pages = {},
title = {Tunable optical spin Hall effect in a liquid crystal microcavity},
volume = {7},
journal = {Light: Science \& Applications},
url = {https://doi.org/10.1038/s41377-018-0076-z},
doi = {10.1038/s41377-018-0076-z}
}

@article{synthetic_Hamiltonians,
author = {Rechcińska, Katarzyna and Król, Mateusz and Mazur, Rafal and Morawiak, Przemysław and Mirek, Rafał and Łempicka-Mirek, Karolina and Bardyszewski, Witold and Matuszewski, Michał and Kula, Przemyslaw and Piecek, Wiktor and Lagoudakis, Pavlos and Pietka, Barbara and Szczytko, Jacek},
year = {2019},
month = {11},
pages = {727-730},
title = {Engineering spin-orbit synthetic Hamiltonians in liquid-crystal optical cavities},
volume = {366},
journal = {Science},
doi = {10.1126/science.aay4182}
}

@article{ulh_morphology,
   author = {Zhixuan Jia and Tejal Pawale and Guillermo I. Guerrero-García and Sid Hashemi and José A. Martínez-González and Xiao Li},
   doi = {10.3390/cryst11040414},
   issn = {20734352},
   issue = {4},
   journal = {Crystals},
   month = {4},
   publisher = {MDPI AG},
   title = {Engineering the uniform lying helical structure in chiral nematic liquid crystals: From morphology transition to dimension control},
   volume = {11},
   year = {2021}
}

@book{kavokin2007microcavities,
  title={Microcavities},
  author={Kavokin, A. and Baumberg, J.J. and Malpuech, G. and Laussy, F.P.},
  isbn={9780199228942},
  lccn={2008297442},
  series={Series on Semiconductor Science and Technology},
  url={https://books.google.pl/books?id=TwK78h12LEUC},
  year={2007},
  publisher={OUP Oxford}
}

@book{Book_topological_insulators,
       author = {{Asb{\'o}th}, J{\'a}nos K. and {Oroszl{\'a}ny}, L{\'a}szl{\'o} and {P{\'a}lyi}, Andr{\'a}s},
        title = "{A Short Course on Topological Insulators}",
         year = {2016},
       volume = {919},
          doi = {10.1007/978-3-319-25607-8},
      publisher ={Springer Cham}
}

@book{Book_topological_matter_quantum_computation,
       author = {Tudor D. Stanescu},
        title = "{Introduction to Topological Quantum Matter and Quantum Computation}",
         year = {2024},
          doi = {10.1201/9781003226048},
        edition = {2nd},
      publisher ={CRC Press.}
}

@article{SSH_PhysRevLett.42.1698_1979,
  title = {Solitons in Polyacetylene},
  author = {Su, W. P. and Schrieffer, J. R. and Heeger, A. J.},
  journal = {Phys. Rev. Lett.},
  volume = {42},
  issue = {25},
  pages = {1698--1701},
  numpages = {0},
  year = {1979},
  month = {Jun},
  publisher = {American Physical Society},
  doi = {10.1103/PhysRevLett.42.1698},
  url = {https://link.aps.org/doi/10.1103/PhysRevLett.42.1698}
}

@article{Muszyński_LasPhotRev2025,
author = {Muszyński, Marcin and Oliwa, Przemysław and Kokhanchik, Pavel and Kapuściński, Piotr and Oton, Eva and Mazur, Rafał and Morawiak, Przemysław and Piecek, Wiktor and Kula, Przemysław and Bardyszewski, Witold and Piętka, Barbara and Bobylev, Daniil and Solnyshkov, Dmitry and Malpuech, Guillaume and Szczytko, Jacek},
title = {Electrically Tunable Spin-Orbit Coupled Photonic Lattice in a Liquid Crystal Microcavity},
journal = {Laser \& Photonics Reviews},
volume = {19},
number = {7},
pages = {2400794},
year = {2025}
}

@article{TaoDu_PhysicaE_2025,
title = {The generalized Su–Schrieffer–Heeger double chains with the chiral and spatial inversion symmetries},
journal = {Physica E: Low-dimensional Systems and Nanostructures},
volume = {172},
pages = {116255},
year = {2025},
issn = {1386-9477},
doi = {https://doi.org/10.1016/j.physe.2025.116255},
url = {https://www.sciencedirect.com/science/article/pii/S1386947725000840},
author = {Tao Du and Yuexun Li and Helin Lu and Hui Zhang}
}

@article{Longhi_OptLett_43_2018,
author = {S. Longhi},
journal = {Opt. Lett.},
number = {19},
pages = {4639--4642},
publisher = {Optica Publishing Group},
title = {Probing one-dimensional topological phases in waveguide lattices with broken chiral symmetry},
volume = {43},
month = {Oct},
year = {2018},
url = {https://opg.optica.org/ol/abstract.cfm?URI=ol-43-19-4639},
doi = {10.1364/OL.43.004639},
}

@article{Zak_PhysRevLett.62.2747_1989,
  title = {Berry's phase for energy bands in solids},
  author = {Zak, J.},
  journal = {Phys. Rev. Lett.},
  volume = {62},
  issue = {23},
  pages = {2747--2750},
  numpages = {0},
  year = {1989},
  month = {Jun},
  publisher = {American Physical Society},
  doi = {10.1103/PhysRevLett.62.2747},
  url = {https://link.aps.org/doi/10.1103/PhysRevLett.62.2747}
}

@article{Lu_NatNanophot_2014,
	author = {Lu, Ling and Joannopoulos, John D. and Solja{\v c}i{\'c}, Marin},
	date = {2014/11/01},
	doi = {10.1038/nphoton.2014.248},
	id = {Lu2014},
	isbn = {1749-4893},
	journal = {Nature Photonics},
	number = {11},
	pages = {821--829},
	title = {Topological photonics},
	url = {https://doi.org/10.1038/nphoton.2014.248},
	volume = {8},
	year = {2014}}

@article{St_Jean_NatNanophot_2017,
	author = {St-Jean, P. and Goblot, V. and Galopin, E. and Lema{\^\i}tre, A. and Ozawa, T. and Le Gratiet, L. and Sagnes, I. and Bloch, J. and Amo, A.},
	date = {2017/10/01},
	doi = {10.1038/s41566-017-0006-2},
	id = {St-Jean2017},
	isbn = {1749-4893},
	journal = {Nature Photonics},
	number = {10},
	pages = {651--656},
	title = {Lasing in topological edge states of a one-dimensionallattice},
	url = {https://doi.org/10.1038/s41566-017-0006-2},
	volume = {11},
	year = {2017}}

@article{Klembt_Nature_2018,
	author = {Klembt, S. and Harder, T. H. and Egorov, O. A. and Winkler, K. and Ge, R. and Bandres, M. A. and Emmerling, M. and Worschech, L. and Liew, T. C. H. and Segev, M. and Schneider, C. and H{\"o}fling, S.},
	date = {2018/10/01},
	doi = {10.1038/s41586-018-0601-5},
	id = {Klembt2018},
	isbn = {1476-4687},
	journal = {Nature},
	number = {7728},
	pages = {552--556},
	title = {Exciton-polariton topological insulator},
	url = {https://doi.org/10.1038/s41586-018-0601-5},
	volume = {562},
	year = {2018}}

@article{Lan_RevPhys_2022,
title = {A brief review of topological photonics in one, two, and three dimensions},
journal = {Reviews in Physics},
volume = {9},
pages = {100076},
year = {2022},
issn = {2405-4283},
doi = {https://doi.org/10.1016/j.revip.2022.100076},
url = {https://www.sciencedirect.com/science/article/pii/S2405428322000077},
author = {Zhihao Lan and Menglin L.N. Chen and Fei Gao and Shuang Zhang and Wei E.I. Sha}
}

@article{Ota_Nanophotonics_2020,
url = {https://doi.org/10.1515/nanoph-2019-0376},
title = {Active topological photonics},
author = {Yasutomo Ota and Kenta Takata and Tomoki Ozawa and Alberto Amo and Zhetao Jia and Boubacar Kante and Masaya Notomi and Yasuhiko Arakawa and Satoshi Iwamoto},
pages = {547--567},
volume = {9},
number = {3},
journal = {Nanophotonics},
doi = {doi:10.1515/nanoph-2019-0376},
year = {2020},
lastchecked = {2026-02-10}
}

@article{A_Yu_Kitaev_2001,
doi = {10.1070/1063-7869/44/10S/S29},
url = {https://doi.org/10.1070/1063-7869/44/10S/S29},
year = {2001},
month = {oct},
publisher = {},
volume = {44},
number = {10S},
pages = {131},
author = {A Yu Kitaev},
title = {Unpaired Majorana fermions in quantum
wires},
journal = {Physics-Uspekhi}
}

@article{Kane1_PhysRevLett.95.146802_2005,
  title = {${Z}_{2}$ Topological Order and the Quantum Spin Hall Effect},
  author = {Kane, C. L. and Mele, E. J.},
  journal = {Phys. Rev. Lett.},
  volume = {95},
  issue = {14},
  pages = {146802},
  numpages = {4},
  year = {2005},
  month = {Sep},
  publisher = {American Physical Society},
  doi = {10.1103/PhysRevLett.95.146802},
  url = {https://link.aps.org/doi/10.1103/PhysRevLett.95.146802}
}

@article{Kane2_PhysRevLett.95.226801_2005,
  title = {Quantum Spin Hall Effect in Graphene},
  author = {Kane, C. L. and Mele, E. J.},
  journal = {Phys. Rev. Lett.},
  volume = {95},
  issue = {22},
  pages = {226801},
  numpages = {4},
  year = {2005},
  month = {Nov},
  publisher = {American Physical Society},
  doi = {10.1103/PhysRevLett.95.226801},
  url = {https://link.aps.org/doi/10.1103/PhysRevLett.95.226801}
}

@article{BHZ_Science.1133734_2006,
author = {B. Andrei Bernevig  and Taylor L. Hughes  and Shou-Cheng Zhang },
title = {Quantum Spin Hall Effect and Topological Phase Transition in HgTe Quantum Wells},
journal = {Science},
volume = {314},
number = {5806},
pages = {1757-1761},
year = {2006},
doi = {10.1126/science.1133734},
URL = {https://www.science.org/doi/abs/10.1126/science.1133734},
eprint = {https://www.science.org/doi/pdf/10.1126/science.1133734}}

@article{Tristan_ACS_Photonics_2021,
author = {Harder, Tristan H. and Sun, Meng and Egorov, Oleg A. and Vakulchyk, Ihor and Beierlein, Johannes and Gagel, Philipp and Emmerling, Monika and Schneider, Christian and Peschel, Ulf and Savenko, Ivan G. and Klembt, Sebastian and H{\"o}fling, Sven},
title = {Coherent Topological Polariton Laser},
journal = {ACS Photonics},
volume = {8},
number = {5},
pages = {1377-1384},
year = {2021},
doi = {10.1021/acsphotonics.0c01958},
URL = {https://doi.org/10.1021/acsphotonics.0c01958}
}

@article{Meier_NatComm_2016,
	author = {Meier, Eric J. and An, Fangzhao Alex and Gadway, Bryce},
	date = {2016/12/23},
	doi = {10.1038/ncomms13986},
	id = {Meier2016},
	isbn = {2041-1723},
	journal = {Nature Communications},
	number = {1},
	pages = {13986},
	title = {Observation of the topological soliton state in the Su--Schrieffer--Heeger model},
	url = {https://doi.org/10.1038/ncomms13986},
	volume = {7},
	year = {2016}}

@article{Jalochowski_ACSNano_2024, 
title={Implementation of the su–schrieffer–heeger model in the self-assembly si–in atomic chains on the si(553)–au surface}, 
volume={18}, 
DOI={10.1021/acsnano.4c00225}, 
number={20}, 
journal={ACS Nano}, 
author={Jałochowski, Mieczysław and Krawiec, Mariusz and Kwapiński, Tomasz}, 
year={2024}, month={May}, 
pages={12861–12869}
}

@article{Cheng_LaserPhotRev_2015,
author = {Cheng, Qingqing and Pan, Yiming and Wang, Qianjin and Li, Tao and Zhu, Shining},
title = {Topologically protected interface mode in plasmonic waveguide arrays},
journal = {Laser \& Photonics Reviews},
volume = {9},
number = {4},
pages = {392-398},
doi = {https://doi.org/10.1002/lpor.201400462},
url = {https://onlinelibrary.wiley.com/doi/abs/10.1002/lpor.201400462},
year = {2015}
}

@article{Whittaker_PhysRevB.99.081402_2019,
  title = {Effect of photonic spin-orbit coupling on the topological edge modes of a Su-Schrieffer-Heeger chain},
  author = {Whittaker, C. E. and Cancellieri, E. and Walker, P. M. and Royall, B. and Tapia Rodriguez, L. E. and Clarke, E. and Whittaker, D. M. and Schomerus, H. and Skolnick, M. S. and Krizhanovskii, D. N.},
  journal = {Phys. Rev. B},
  volume = {99},
  issue = {8},
  pages = {081402},
  numpages = {5},
  year = {2019},
  month = {Feb},
  publisher = {American Physical Society},
  doi = {10.1103/PhysRevB.99.081402},
  url = {https://link.aps.org/doi/10.1103/PhysRevB.99.081402}
}

@misc{rufo2025domainwallcontroltopological,
      title={Domain Wall Control of Topological Qubits in the Kitaev SSH Chain}, 
      author={Griffith Rufo and Sabrina Rufo and Heron Caldas and Rosiane de Freitas},
      year={2025},
      eprint={2512.09693},
      archivePrefix={arXiv},
      primaryClass={cond-mat.mes-hall},
      url={https://arxiv.org/abs/2512.09693}, 
}

@article{LiuPhysRevB.104.085302_2021,
  title = {Topological states and interplay between spin-orbit and Zeeman interactions in a spinful Su-Schrieffer-Heeger nanowire},
  author = {Liu, Zhi-Hai and Entin-Wohlman, O. and Aharony, A. and You, J. Q. and Xu, H. Q.},
  journal = {Phys. Rev. B},
  volume = {104},
  issue = {8},
  pages = {085302},
  numpages = {11},
  year = {2021},
  month = {Aug},
  publisher = {American Physical Society},
  doi = {10.1103/PhysRevB.104.085302},
  url = {https://link.aps.org/doi/10.1103/PhysRevB.104.085302}
}

@article{Pham_PhysRevB.105.125418_2022,
  title = {Topological states in dimerized quantum-dot chains created by atom manipulation},
  author = {Pham, Van Dong and Pan, Yi and Erwin, Steven C. and von Oppen, Felix and Kanisawa, Kiyoshi and F\"olsch, Stefan},
  journal = {Phys. Rev. B},
  volume = {105},
  issue = {12},
  pages = {125418},
  numpages = {9},
  year = {2022},
  month = {Mar},
  publisher = {American Physical Society},
  doi = {10.1103/PhysRevB.105.125418},
  url = {https://link.aps.org/doi/10.1103/PhysRevB.105.125418}
}

@article{Zhao_PhysRevB.110.235106_2024,
  title = {Interplay of topology and interaction in an exactly solvable SSH-BCS-Hubbard chain},
  author = {Zhao, Yuejiu and Miao, Jian-Jian and Zhang, Fu-Chun},
  journal = {Phys. Rev. B},
  volume = {110},
  issue = {23},
  pages = {235106},
  numpages = {11},
  year = {2024},
  month = {Dec},
  publisher = {American Physical Society},
  doi = {10.1103/PhysRevB.110.235106},
  url = {https://link.aps.org/doi/10.1103/PhysRevB.110.235106}
}

@article{Chang_SciRep_2025,
	author = {Chang, Pei-Jie and Pi, Jinghui and Zheng, Muxi and Lei, Yu-Ting and Pan, Xingbo and Ruan, Dong and Long, Gui-Lu},
	date = {2025/05/23},
	doi = {10.1038/s41598-025-02284-5},
	id = {Chang2025},
	isbn = {2045-2322},
	journal = {Scientific Reports},
	number = {1},
	pages = {18023},
	title = {Topological phases of extended Su-Schrieffer-Heeger-Hubbard model},
	url = {https://doi.org/10.1038/s41598-025-02284-5},
	volume = {15},
	year = {2025}}

@misc{wang2026defectengineeringspincenters,
      title={Defect engineering spin centers in interacting many-body Su-Schrieffer-Heeger chains}, 
      author={Lin Wang and Thomas Luu and Ulf-G. Meißner},
      year={2026},
      eprint={2507.18806},
      archivePrefix={arXiv},
      primaryClass={cond-mat.str-el},
      doi={https://doi.org/10.1103/n1h1-4vtr},
      url={https://arxiv.org/abs/2507.18806}, 
}

@article{Tamura_PhysRevB.101.214507_2020,
  title = {Bulk odd-frequency pairing in the superconducting Su-Schrieffer-Heeger model},
  author = {Tamura, Shun and Nakosai, Sho and Black-Schaffer, Annica M. and Tanaka, Yukio and Cayao, Jorge},
  journal = {Phys. Rev. B},
  volume = {101},
  issue = {21},
  pages = {214507},
  numpages = {14},
  year = {2020},
  month = {Jun},
  publisher = {American Physical Society},
  doi = {10.1103/PhysRevB.101.214507},
  url = {https://link.aps.org/doi/10.1103/PhysRevB.101.214507}
}

@article{Ryu_NewJourPhys_2010,
doi = {10.1088/1367-2630/12/6/065010},
url = {https://doi.org/10.1088/1367-2630/12/6/065010},
year = {2010},
month = {jun},
publisher = {},
volume = {12},
number = {6},
pages = {065010},
author = {Ryu, Shinsei and Schnyder, Andreas P and Furusaki, Akira and Ludwig, Andreas W W},
title = {Topological insulators and superconductors: tenfold way and dimensional hierarchy},
journal = {New Journal of Physics}
}

@article{Delplace_PhysRevB.84.195452_2011,
  title = {Zak phase and the existence of edge states in graphene},
  author = {Delplace, P. and Ullmo, D. and Montambaux, G.},
  journal = {Phys. Rev. B},
  volume = {84},
  issue = {19},
  pages = {195452},
  numpages = {13},
  year = {2011},
  month = {Nov},
  publisher = {American Physical Society},
  doi = {10.1103/PhysRevB.84.195452},
  url = {https://link.aps.org/doi/10.1103/PhysRevB.84.195452}
}

@article{merons,
author = {Mateusz Kr\'{o}l and Helgi Sigurdsson and Katarzyna Rechci\'{n}ska and Przemys{\l}aw Oliwa and Krzystof Tyszka and Witold Bardyszewski and Andrzej Opala and Micha{\l} Matuszewski and Przemys{\l}aw Morawiak and Rafa{\l} Mazur and Wiktor Piecek and Przemys{\l}aw Kula and Pavlos G. Lagoudakis and Barbara Piętka and Jacek Szczytko},
journal = {Optica},
number = {2},
pages = {255--261},
publisher = {Optica Publishing Group},
title = {Observation of second-order meron polarization textures in optical microcavities},
volume = {8},
month = {Feb},
year = {2021},
url = {https://opg.optica.org/optica/abstract.cfm?URI=optica-8-2-255},
doi = {10.1364/OPTICA.414891},
}

@article{psh,
  title = {Realizing Optical Persistent Spin Helix and Stern-Gerlach Deflection in an Anisotropic Liquid Crystal Microcavity},
  author = {Kr\'ol, Mateusz and Rechci\ifmmode \acute{n}\else \'{n}\fi{}ska, Katarzyna and Sigurdsson, Helgi and Oliwa, Przemys\l{}aw and Mazur, Rafa\l{} and Morawiak, Przemys\l{}aw and Piecek, Wiktor and Kula, Przemys\l{}aw and Lagoudakis, Pavlos G. and Matuszewski, Micha\l{} and Bardyszewski, Witold and Pi\ifmmode \mbox{\k{e}}\else \k{e}\fi{}tka, Barbara and Szczytko, Jacek},
  journal = {Phys. Rev. Lett.},
  volume = {127},
  issue = {19},
  pages = {190401},
  numpages = {7},
  year = {2021},
  month = {Nov},
  publisher = {American Physical Society},
  doi = {10.1103/PhysRevLett.127.190401},
  url = {https://link.aps.org/doi/10.1103/PhysRevLett.127.190401}
}

@article{pol_singularities_lcmc,
author = {Oliwa, Przemysław and Kapuściński, Piotr and Popławska, Maria and Muszyński, Marcin and Król, Mateusz and Morawiak, Przemysław and Mazur, Rafał and Piecek, Wiktor and Kula, Przemysław and Bardyszewski, Witold and Piętka, Barbara and Sigurðsson, Helgi and Szczytko, Jacek},
title = {Electrically Tunable Momentum Space Polarization Singularities in Liquid Crystal Microcavities},
journal = {Advanced Science},
volume = {12},
number = {23},
pages = {2500060},
doi = {https://doi.org/10.1002/advs.202500060},
year = {2025}
}

@article{gaussian_traps,
author = {Rafa{\l} Mirek and Pavel Kokhanchik and Darius Urbonas and Ioannis Georgakilas and Marcin Muszy\'{n}ski and Piotr Kapu\'{s}ci\'{n}ski and Przemys{\l}aw Oliwa and Barbara Piętka and Jacek Szczytko and Michael Forster and Ullrich Scherf and Przemys{\l}aw Morawiak and Wiktor Piecek and Przemys{\l}aw Kula and Dmitry Solnyshkov and Guillaume Malpuech and Rainer F. Mahrt and Thilo St\"{o}ferle},
journal = {Optica},
number = {9},
pages = {1548--1552},
publisher = {Optica Publishing Group},
title = {In situ tunneling control in photonic potentials by Rashba--Dresselhaus spin--orbit coupling},
volume = {12},
month = {Sep},
year = {2025},
url = {https://opg.optica.org/optica/abstract.cfm?URI=optica-12-9-1548},
doi = {10.1364/OPTICA.555447},
}

@article{
torony,
author = {Marcin Muszyński  and Daniil Bobylev  and Piotr Kapuściński  and Przemysław Oliwa  and Joanna Mędrzycka  and Eva Oton  and Rafał Mazur  and Przemysław Morawiak  and Wiktor Piecek  and Przemysław Kula  and Dmitry Solnyshkov  and Guillaume Malpuech  and Jacek Szczytko },
title = {Ground-state orbital angular momentum lasing from liquid crystal torons embedded in a microcavity},
journal = {Science Advances},
volume = {12},
number = {11},
pages = {eaeb6167},
year = {2026},
doi = {10.1126/sciadv.aeb6167},
URL = {https://www.science.org/doi/abs/10.1126/sciadv.aeb6167},
eprint = {https://www.science.org/doi/pdf/10.1126/sciadv.aeb6167}}

@book{IbachLuch,
  title={Solid-State Physics: An Introduction to Principles of Materials Science},
  author={Ibach, H. and L{\"u}th, H.},
  isbn={9783540938040},
  lccn={2009926968},
  series={Physics and Astronomy},
  url={https://books.google.pl/books?id=qjxv68JFe3gC},
  year={2009},
  publisher={Springer Berlin Heidelberg}
}
\end{document}